\documentclass[11pt]{article}
\usepackage{graphicx}
\usepackage{epsf}  

\newcommand{\BABARPubYear}    {03}

\newcommand{\BABARConfNumber} {010}
\newcommand{\SLACPubNumber} {9693}
\newcommand{\LANLNumber} {0303040}

\input pubboard/babarsym
\usepackage{BAD}

\providecommand{\OnEvtOmk}{\mbox{$16729$}}
\providecommand{\OnEvtOmp}{\mbox{$30563$}}

\providecommand{\YOmkInt}{\mbox{$87$}}

\providecommand{\YOmpInt}{\mbox{$101$}}

\providecommand{\YOmksInt}{\mbox{$33$}}

\long\def\inst#1{\par\nobreak\kern 4pt\nobreak
    {\it #1}\par\vskip 10pt plus 3pt minus 3pt}

\begin{document}
{\pagestyle{empty}

\begin{flushleft}
\end{flushleft}
\begin{flushright}
\babar-CONF-\BABARPubYear/\BABARConfNumber \\
SLAC-PUB-\SLACPubNumber \\
hep-ex/\LANLNumber \\
March 2003 \\
\end{flushright}

\par\vskip 5cm  

\begin{center}
\Large \bf\boldmath 
Observation of $B$ Meson Decays to $\omega\pi^+$, $\omega\Kp$, and $\omega\Kz$
\end{center}
\bigskip

\begin{center}
\large The \babar\ Collaboration\\
\mbox{ }\\
\today
\end{center}
\bigskip \bigskip

\begin{center}
\large \bf Abstract
\end{center}
We present preliminary measurements of $B$ meson decays to \omegapi, \omegaKp,
and \omegaKz.  The data were recorded with the \babar\ detector and
correspond to $88.9\times 10^6$ \BB\ pairs produced in \epem\
annihilation at the \UfourS\ resonance.
We find statistically significant signals for all three channels:
$\Bomegapi = \Romegapi$, $\BomegaKp = \RomegaKp$,
and $\BomegaKz = \RomegaKz$.
We also measure time-integrated charge asymmetries $\acp(\omegapi)=\Aomegapi$ and
$\acp(\omegaKp)=\AomegaKp$.

\vfill
\begin{center}
Presented at the XXXVIII$^{th}$ Rencontres de Moriond on\\
QCD and High Energy Hadronic Interactions, \\
3/22---3/29/2003, Les Arcs, Savoie, France
\end{center}

\vspace{1.0cm}
\begin{center}
{\em Stanford Linear Accelerator Center, Stanford University, 
Stanford, CA 94309} \\ \vspace{0.1cm}\hrule\vspace{0.1cm}
Work supported in part by Department of Energy contract DE-AC03-76SF00515.
\end{center}

\newpage
} 

\begin{center}
\small

The \babar\ Collaboration,
\bigskip

%
B.~Aubert,
R.~Barate,
D.~Boutigny,
J.-M.~Gaillard,
A.~Hicheur,
Y.~Karyotakis,
J.~P.~Lees,
P.~Robbe,
V.~Tisserand,
A.~Zghiche
\inst{Laboratoire de Physique des Particules, F-74941 Annecy-le-Vieux, France }
% ...

\end{center}\newpage

\section{Introduction}
\label{sec:intro}

We report the results of searches for $B$ decays to the charmless
final states $\omega\pi^+$, $\omega\Kp$, and $\omega\Kz$.  We reconstruct the
$\omega$ mesons via the dominant decay mode $\omega\ra\pi^+\pi^-\pi^0$,
and \Kz\ via $\Kz\ra\KS\ra\pi^+\pi^-$.
For the charged modes we also measure the direct \CP-violating
time-integrated charge asymmetry $\acp=(\Gamma^--\Gamma^+)/(\Gamma^-+\Gamma^+)$,
where $\Gamma^\pm\equiv\Gamma(B^\pm\ra\omega h^\pm)$.

Table~\ref{tab:OldResults} summarizes the current knowledge of these decays,
coming from measurements by CLEO~\cite{CLEOomega}, \babar\ 
\cite{BABARomega, BloomSSI} and Belle~\cite{BELLEomega}.  
CLEO and \babar\ find significant signals for the \omegapi\ channel.  The
\omegaKp\ decay has an interesting history.  It was originally seen by
CLEO but a re-analysis with the full CLEO dataset could not confirm the earlier 
large branching fraction measurement.  \babar\ confirmed the smaller result
of the CLEO re-analysis (see Table~\ref{tab:OldResults}) but Belle has now 
published a significant observation with a relatively large branching 
fraction.  A more precise measurement would help settle the situation.
\babar\ reported a significant signal for the \omegaKz\ decay 
at conferences in summer 2002.

The early indications from CLEO and now
Belle of a branching fraction for \omegaKp\ of $\gsim10\times10^{-6}$ were
hard to accommodate theoretically.  The \omegaKp\ decay is an interesting 
penguin-dominated decay with cancellations between primary Wilson
coefficients.  The \omegapi\ decay is expected to be dominated by the
external and color-supressed tree diagrams.  Typical calculations 
found branching fractions for both $\omega$ decays of a few times $10^{-6}$ 
\cite{ali,cheng,dutta}, although parameter tuning was able to accommodate
$\gsim10\times10^{-6}$ with difficulty.  The decay \omegaKz\ is expected
to have a comparable branching fraction to that of \omegaKp.  However 
the presence of a (suppressed) tree diagram in the charged decay would
reduce the rate 
than for the neutral decay if there is substantial destructive
tree-penguin interference for \omegaKp.  The theoretical situation has
changed little recently.  There are few calculations for these modes
from the QCD factorization or perturbative QCD groups.  One recent
calculation of pseudoscalar-vector modes with QCD factorization \cite{aleksan}
finds predicted branching fractions of $\lsim5\times10^{-6}$ for all three
decays.

\begin{table}[htb]
\caption{Summary of branching fraction results for $B$ decays to $\omega$
    mesons from CLEO \cite{CLEOomega}, previous \babar\ measurements 
    \cite{BABARomega, BloomSSI}, Belle \cite{BELLEomega}, and the
    present analysis.  The results for all fits are given as well as a 90\%
    confidence level upper limit if the measured yield is not judged to be 
    significant.  The signal yields and efficiencies ($\epsilon$) are also given.
}
\label{tab:OldResults}
\begin{center}
\vspace*{0.5cm}
\hspace*{-0.5cm}
\begin{tabular}{l|ccccccc}
    \dbline
Expt.    &\# \BB\ ($\times10^{6}$)&Fit \calB$(\times10^{-6})$&UL \calB$(\times10^{-6})$&Signif. ($\sigma$)&Signal yield&$\epsilon$ (\%) \\
    \sgline
$\omegaKp$ & & & & & & \\
~~CLEO     &10& $3.2^{+2.4}_{-1.9}\pm0.8$&7.9&2.1& $7.9^{+6.0}_{-4.7}$& 26 \\
~~\babar   &23& $1.4^{+1.3}_{-1.0}\pm0.3$&3.3&1.6& $6.4^{+5.6}_{-4.4}$& 19 \\
~~Belle    &32& $9.2^{+2.6}_{-2.3}\pm1.0$& - &6.0&$18.9^{+5.4}_{-4.7}$& 6.0 \\
This result&89&       \romegaKp          & - &8.9& $87\pm15$          & 18 \\
    \sgline
$\omegapi$ & & & & & & \\
~~CLEO     &10&$11.3^{+3.3}_{-2.9}\pm0.8$& - &6.2&$28.5^{+8.2}_{-7.3}$& 26 \\
~~\babar   &23& $6.6^{+2.1}_{-1.8}\pm0.7$& - &5.1&$28^{+9}_{-8}$      & 19 \\
~~Belle    &32& $4.2^{+2.0}_{-1.8}\pm0.5$&8.1&3.3&$10.4^{+4.7}_{-4.3}$&7.7 \\
This result&89&      \romegapi           & - &8.4&$101\pm18$          & 19 \\
    \sgline
$\omegaKz$ & & & & & & \\
~~CLEO     &10&$10.0^{+5.4}_{-4.2}\pm1.4$&21 &3.9&$7.0^{+3.0}_{-2.9}$ & 7.4 \\
~~\babar   &62& $5.9^{+1.7}_{-1.5}\pm0.9$& - &6.6&$27^{+8}_{-7}$      & 7.4 \\
This result&89&      \romegaKz           & - &7.5&$33^{+9}_{-8}$      & 7.0 \\
    \dbline
\end{tabular}
\end{center}
\end{table}

\section{Detector and Data} \label{sec:detector}

The results presented in this paper are based on data collected
in 1999--2002 with the \babar\ detector~\cite{BABARNIM}
at the PEP-II asymmetric-energy $e^+e^-$ collider~\cite{pep}
located at the Stanford Linear Accelerator Center.  An integrated
luminosity of 81.9~fb$^{-1}$, corresponding to 
88.9 million \BB\ pairs, was recorded at the $\Upsilon (4S)$
resonance
(``on-resonance'', center-of-mass energy $\sqrt{s}=10.58\ \gev$).
An additional 9.6~fb$^{-1}$ were taken about 40~MeV below
this energy (``off-resonance'') for the study of continuum backgrounds in
which a light or charm quark pair is produced instead of an \UfourS.

The asymmetric beam configuration in the laboratory frame
provides a boost of $\langle\beta\gamma\rangle = 0.56$ to the $\Upsilon(4S)$.
Charged particles are detected and their momenta measured by a
combination of a silicon vertex tracker (SVT), consisting of five layers
of double-sided detectors, and a 40-layer central drift chamber,
both operating in the 1.5~T magnetic field of a solenoid. Photons and
electrons are detected by a CsI(Tl) electromagnetic calorimeter (EMC).

Charged particle identification (PID) is provided primarily
by an internally reflecting ring-imaging 
Cherenkov detector (DIRC) covering the central region, though the average 
energy loss ($dE/dx$) in the tracking devices is also used for the pion
daughters from the $\omega$ decay.
From the Cherenkov angle, a $K$--$\pi$ separation greater than 4 standard
deviations ($\sigma$) is 
achieved for tracks with momenta below $3\ \gevc$, decreasing to 
$2.5\,\sigma$ at the highest momenta in the final states considered here
\cite{KpiPRL}.

\section{Event Selection} 
\label{sec:presel}

Monte Carlo (MC) simulations \cite{geant}\ of the target decay modes and
of continuum and \BB\ backgrounds are used to establish the event selection
criteria.  The selection is designed to achieve high efficiency and
retain sidebands sufficient to characterize the background for
subsequent fitting.  

From photons with energy $E_\gamma>50$ MeV, we select $\piz$ candidates by 
requiring the invariant mass to satisfy $120<m_{\gamma\gamma}< 150$ MeV.
Candidate $\omega$ mesons must satisfy $735<m_{\pi\pi\pi}<825$ MeV.
For \kzs\ candidates we require $488<m_{\pi\pi}< 508$ MeV,
the three-dimensional flight
distance from the event primary vertex $>2$ mm, and the
two-dimensional angle between flight and momentum vectors $<40$ mrad.

We make several particle identification requirements to ensure the
identity of the signal pions and kaons.
Daughter tracks of $\omega$ candidates must have DIRC, $dE/dx$, and EMC
responses consistent with pions.  For the \omegaKp\ decay, the
prompt charged track must have an associated DIRC Cherenkov angle between
$-5\,\sigma$ and $+2\,\sigma$ of the value expected for a kaon.  
For \omegapi, the DIRC Cherenkov angle must be between
$-2\,\sigma$ and $+5\,\sigma$ of the value expected for a pion.

A $B$ meson candidate is characterized kinematically by the energy-substituted 
mass $\mes = \sqrt{(\half s + \pvec_0\cdot \pvec_B)^2/E_0^2 - \pvec_B^2}$ and
energy difference $\DE = E_B^*-\half\sqrt{s}$, where the subscripts 0 and
$B$ refer to the initial \UfourS\ and the $B$ candidate, respectively,
and the asterisk denotes the \UfourS\ frame. We require $|\DE|\le0.2$ GeV and
$5.2\le\mes\le5.29\ \gev$.
The resolutions on these quantities are
about 30 MeV and $3.0\ \mev$, respectively.

\subsection{Tau, QED, and continuum background}

To discriminate against tau-pair and two-photon background, we require
that the event contain at least five (four) 
charged tracks for neutral (charged) $B$ pairs.  
To reject continuum background, we define
an angle $\theta_T$ between the thrust axes of the $B$ candidate and of
the rest of the tracks and neutral clusters in the event, calculated in
the center-of-mass frame.  The distribution of $|\costhr|$ is
sharply peaked near 1.0 for combinations drawn from jet-like \qqbar\
pairs, and nearly uniform for the isotropic $B$ meson decays; we require
$|\costhr|<0.8$ for the charged modes and $|\costhr|<0.8$ for the 
lower-background \omegaKz\ decay.  A second $B$ candidate satisfying the selection
criteria is found in about 10--20\% of the events.  In this case the ``best''
combination is chosen as the one closest to the nominal $\omega$ mass.

\subsection{\boldmath \BB\ background}

We use MC simulations of \BzBzb\ and \BpBm\ pair production and decay to
look for possible \BB\ backgrounds.  
From these studies we find that
\BB\ background is small for all three decay channels.  The uncertainty
is limited by the statistical errors of these studies and is included in
the systematic errors (see \S\ref{sec:syst}).

\section{Maximum Likelihood Fit}\label{sec:mlfit}

We use an unbinned, multivariate maximum-likelihood fit to extract signal yields 
for our modes.  With the cuts in \S\ref{sec:presel}, candidates are selected to 
match the kinematic structure of the desired decay chain.

\subsection{Likelihood Function} \label{sec:like}

We incorporate several uncorrelated variables for the kinematics of the
$B$ decay chain and a Fisher discriminant \xf\ for the $B$ production and
energy flow.  Thus the input observables are \DE, \mes, $m_{\pi\pi\pi}$,
$\hel\equiv |\cos{\theta_H}|$, and a Fisher discriminant \xf.
The helicity angle $\theta_H$ is defined as the angle, measured in the
$\omega$ rest frame, between the normal to the $\omega$ decay plane and the 
flight direction of the $\omega$.  
The Fisher discriminant \cite{CLEO-fisher}\ combines four
variables: the angles with respect to the beam axis, in the \UfourS\
frame, of the $B$ momentum and $B$ thrust axis, and
the zeroth and second angular moments $L_{0,2}$ of the energy flow
about the $B$ thrust axis.  The moments are defined by
\begin{equation}
  L_j = \sum_i p_i\times\left|\cos\theta_i\right|^j,
\end{equation}
where $\theta_i$ is the angle with respect to the $B$ thrust axis of
track or neutral cluster $i$, $p_i$ is its momentum, and the sum
excludes the $B$ candidate.

Since we measure the correlations 
among the observables in the data to be small, we take the probability density 
function (PDF) for each event to be a product of the PDFs for the separate
observables.  We define two hypotheses $j$, where $j$ represents signal or
continuum background.  The product PDF (to be evaluated with the
observable set for event $i$) is then given by

\begin{equation}
{\cal P}^i_{j} =  {\cal P}_j (\mes) \cdot {\cal  P}_j (\DE) \cdot
 { \cal P}_j(\xf) \cdot {\cal P}_j (m_{\pi\pi\pi})\cdot {\cal P}_j (\hel)\ .
\end{equation}

\noindent The likelihood function for each decay chain is

\begin{equation}
\calL=\frac{\exp{(-\sum_j Y_{j})}}{N!}\prod_i^{N}\sum_j Y_{j}{\cal P}^i_{j}\,,
\end{equation}

\noindent where $Y_{j}$ is the yield of events of hypothesis $j$ found by 
maximizing \calL\
and $N$ is the number of events in the sample.  The first factor takes 
into account the Poisson fluctuations in the total number of events. 

\subsection{Fit Parameters}

The determination of PDF parameters for the likelihood fit is accomplished 
with use of Monte Carlo (MC) simulation for the signal and on-peak data from a 
\mes-\DE\ sideband for the continuum background.  Several of the principle
background parameters are allowed to float in the final fit. 

Peaking distributions (signal
masses, \DE, and \xf) are parameterized with Gaussian functions, with a
second or third Gaussian or asymmetric width as required for good fits to these
samples.  Because these are pseudoscalar--vector decays of the $B$, the
signal helicity-angle distribution is proportional to $\hel^2$.
Slowly varying distributions (mass, energy or helicity-angle distributions
for combinatoric background) have low order polynomial
shapes, with the peaking and combinatoric $\omega$ mass components each
having their own $\hel$ shape.  The combinatoric background in \mes\ is 
described by a phase-space-motivated empirical function \cite{argus}.
The background Fisher PDF contains a second Gaussian component wide enough
to ensure the PDF is not excessively small anywhere in its domain.

Control samples of $B$ decays to charm final states of similar topology
are used to test the quality of the MC simulation for variables describing
$B$ decay kinematics.  
Where the control data samples reveal differences from MC in mass or energy
resolution, we scale the resolution used in the likelihood fits. 

The efficiency bias of the likelihood fit is determined from
simulated data samples. The \qq\ background in these samples is generated 
from the continuum PDF shapes.  A small number of signal events from the
detailed MC simulation are added to create a sample with the same
size as the data.  The bias in the fit efficiency is determined from 
the mean of the fit yield for 500 such simulated samples.

\section{Fit Results}

By generating and fitting samples with both signal 
and continuum background events generated from the PDFs, we verify that our 
fitting procedure is functioning properly.  We find that the minimum $-\ln{\cal L}$ 
value in the on-resonance sample lies well within the $-\ln{\cal L}$ 
distribution from these simulated samples.

In Table~\ref{tab:res_omegahks} we show the results of the fits for
off-peak and on-peak data.  
Shown for each decay mode are the number of events that were fit, the
signal yield, the fully corrected efficiency and product branching
fraction (for the $\omega$ and \KS\ decays), the measured branching fraction, 
and the statistical significance of the result.  We also give the branching 
fraction after correction for a crossfeed between the two charged channels.  
The $K/\pi$ misidentification rate of $9\pm2\%$ 
is found in studies with kaon and pion samples tagged kinematically from the 
decay $D^{*+}\ra\pi^+D^0$ with $D^0\ra K^-\pi^+$.
The statistical error on the number of events
is taken as the change in the central value when the quantity
$-2\ln{\cal L}$ changes by one unit. The statistical significance is
taken as the square root of the difference between the value of
$-2\ln{\cal L}$ for zero signal and the value at its minimum.  We also
give the charge asymmetry \acp\ for the charged modes for both signal 
and \qq\ background.

%
%

\begin{table}[tbp]
\vspace{-0.2cm}
\caption{
    Fit values for \fomegaKp, \fomegapi, and \fomegaKz.  The corrected
$\calB$ for the charged modes is the branching fraction after correcting
for crossfeed from one charged mode into the other.
}
\label{tab:res_omegahks}
\hspace*{-0.5cm}
\begin{center}
\begin{tabular}{lccc}
    \dbline
    Fit quantity &\fomegaKp      &\fomegapi      &\fomegaKz      \\
    \sgline
    Fit sample size &               &               &               \\
    ~~On-resonance  & \OnEvtOmk     & \OnEvtOmp         & 9563  \\
    ~~Off-resonance & 1900          & 3490          & 972   \\
    Signal yield    &               &               &       \\
    ~~On-res data	&
$\YOmkInt\pm15$&$\YOmpInt\pm18$&$\YOmksInt^{+9}_{-8}$ \\
    ~~Off-res data  & $0.0^{+2.7}_{-0.0}$ & $0.0^{+3.6}_{-0.0}$
                    & $0.0^{+0.9}_{-0.0}$ \\
    Combinations/event  & 1.13      & 1.15          & 1.15  \\
    Selection $\epsilon$ (\%)  & 20.0      & 21.9          & 23.1  \\
    $\prod\calB_i$ (\%)& 88.8       &   88.8        & 30.5  \\
    \corrEffB       & 17.8          & 19.4          & 7.0  \\ 
    Stat. significance ($\sigma$)  & 8.9   & 8.4            & 7.5    \\ 
    \sgline			 
    \bfemsix    & $5.5\pm1.0\pm0.4$               & $5.9\pm1.0\pm0.5$
                &  \romegaKz  \\
    \sgline
    Corrected \bfemsix    & \romegaKp         & \romegapi   \\
    \sgline
Signal \acp     & \AomegaKp        &  \Aomegapi       & --- \\
Background \acp & $-0.005\pm0.008$ & $-0.012\pm0.006$ & --- \\
    \dbline
\end{tabular}
\end{center}
\end{table}

\begin{figure}[!htb]
\vspace{0.5cm}
 \includegraphics[angle=0,width=\linewidth]{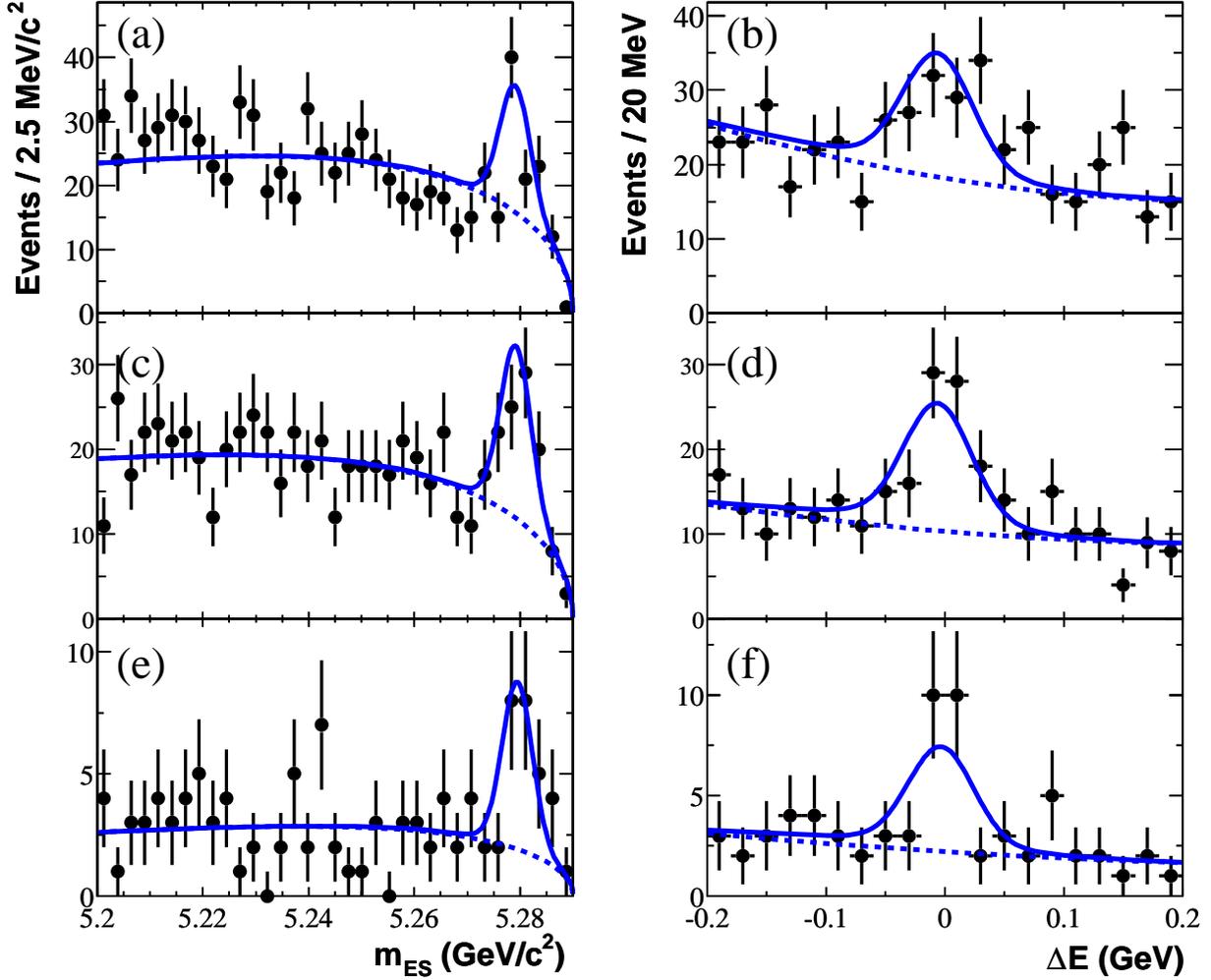}
 \caption{\label{fig:projMbDE}
 The $B$ candidate \mes\ and \DE\ for \omegapi\ (a, b), \omegaKp\ (c, d), and
\omegaKz\ (e, f).
Points with errors represent data passing a cut on a likelihood ratio
calculated without the quantity that is shown in the plots.  
The solid curves show the projected fit functions
and dashed curves the background functions.
  }
\end{figure}

In Fig.\ \ref{fig:projMbDE}\ we show projections of \mes\ and \DE\ made by
selecting events with signal likelihood (computed without the variable
shown in the figure) exceeding a mode-dependent threshold that optimizes the
expected sensitivity.

\section{Systematic Uncertainties}
\label{sec:syst}

Most of the systematic errors on yields that arise from uncertainties in the 
values of the PDF parameters have already been incorporated into the overall
statistical error, because their background parameters are free in the
fit.  We determine the sensitivity to parameters of the signal PDF
components by varying these within their uncertainties.  The results are
shown in the first row of Table \ref{tab:systtab}.  This is the only
additive systematic error.

The uncertainty in our knowledge of the efficiency is found from
auxiliary studies to be 0.8$N_t$\%, 2.5$N_\gamma$\%, and 3\%\ for a
\KS\ decay, where $N_t$ and $N_\gamma$ are the number of signal tracks
and photons, respectively.  We estimate the uncertainty in the number of 
produced \BB\ pairs to be 1.1\%.  The 
systematic bias from the fitter itself (1--2\%) is estimated from fits of
simulated samples with varying background populations.  Published world
averages \cite{PDG2002}\ provide the $B$ daughter branching fraction
uncertainties.  We account for systematic effects in \costhr\ (1\%) and in the 
PID requirements (0.5\%).  Values for each of
these contributions are given in Table \ref{tab:systtab}. 

A study of the charge asymmetry as a function of momentum for all tracks in 
hadronic events bounds the tracking efficiency component of a charge-asymmetry 
bias to be below 1\%.  $D^*$-tagged $D\ra K\pi$ and $B$ samples provide
additional crosschecks that the bias is negligible.
We assign a systematic uncertainty for \acp\ of 1.1\% based on the tracking
study and a small PID contribution determined from the $D^*$ studies.

\begin{table}[htbp]
\caption{
    Estimates of systematic errors (in percent) for the
    \fomegaKp, \fomegapi and \fomegaKz\ branching fractions.
}
\label{tab:systtab}
\begin{center}
\begin{tabular}{l|ccc}
    \dbline
    Quantity    & \fomegaKp & \fomegapi &\fomegaKz \\
    \sgline
    Fit yield               & 2.7       & 3.1   & 3.2   \\
    Fit efficiency/bias     & 2.7	& 3.9 	& 2.1 	\\
    Track multiplicity      & 1.0       & 1.0   & 1.0 	\\
    Tracking eff/qual       & 2.4	& 2.4	& 3.7  	\\
    \piz / $\gamma$ eff     & 5.0       & 5.0   & 5.0	\\
    \KS\ efficiency         & ---	& ---	& 2.9  	\\
    Number \BB\             & 1.1	& 1.1	& 1.1 	\\
    Branching fractions     & 1.0       & 1.0   & 1.0 	\\
    MC statistics           & 1.0	& 1.0	& 1.0 	\\
    \costhr                 & 1.0       & 1.0   & 1.0 	\\
    PID                     & 1.4	& 1.4	& 1.0  	\\
    \BB\ Background         & 1.1	& 1.0	& 3.0  	\\
    \sgline
    Total                   & 7.3       & 8.0   & 8.8  \\
    \dbline
\end{tabular}
\end{center}
\end{table}

\section{Conclusion}

We report preliminary measurements of branching fractions for the decays 
\omegapi, \omegaKp, and \omegaKz.  
We find statistically significant signals for all three decays and
measure the following branching fractions:
\begin{eqnarray*}
\Bomegapi &=& \Romegapi, \\
\BomegaKp &=& \RomegaKp, \\
\BomegaKz &=& \RomegaKz.
\end{eqnarray*}
These results supersede the previous \babar\ measurements 
\cite{BABARomega, BloomSSI}.  The result for \BomegaKp\ is larger than
the previous measurement with one quarter the luminosity.  From studies
of the old and new measurements, we conclude that the difference is a 
statistical fluctuation.

These results are much more precise than previous results and in 
good agreement with theoretical expectations.  A branching fraction for 
\omegaKp\ near $10\times10^{-6}$, as reported first by CLEO and more
recently by Belle,
is now definitively ruled out.  In addition, we measure the charge
asymmetries $\acp(\omegapi)=\Aomegapi$ and $\acp(\omegaKp)=\AomegaKp$.
These values are consistent with being small as generally expected for 
these decays \cite{AKL} though both experimental and theoretical 
uncertainties are still large.

\section{Acknowledgments}
\label{sec:Acknowledgments}

We are grateful for the 
extraordinary contributions of our \pep2\ colleagues in
achieving the excellent luminosity and machine conditions
that have made this work possible.
The success of this project also relies critically on the 
expertise and dedication of the computing organizations that 
support \babar.
The collaborating institutions wish to thank 
SLAC for its support and the kind hospitality extended to them. 
This work is supported by the
US Department of Energy
and National Science Foundation, the
Natural Sciences and Engineering Research Council (Canada),
Institute of High Energy Physics (China), the
Commissariat \`a l'Energie Atomique and
Institut National de Physique Nucl\'eaire et de Physique des Particules
(France), the
Bundesministerium f\"ur Bildung und Forschung and
Deutsche Forschungsgemeinschaft
(Germany), the
Istituto Nazionale di Fisica Nucleare (Italy),
the Foundation for Fundamental Research on Matter (The Netherlands),
the Research Council of Norway, the
Ministry of Science and Technology of the Russian Federation, and the
Particle Physics and Astronomy Research Council (United Kingdom). 
Individuals have received support from 
the A. P. Sloan Foundation, 
the Research Corporation,
and the Alexander von Humboldt Foundation.

\end{document}